\documentclass[journal=nalefd,manuscript=article]{achemso}

\usepackage{graphicx}

\usepackage[version=3]{mhchem}   % Formula subscripts using \ce{}

\author{Simon Abay}
\email{abay@chalmers.se}
\affiliation[Chalmers University of Technology]
{Department of Microtechnology and Nanoscience (MC2), Chalmers University of Technology, SE-412 96 G\"oteborg, Sweden}
\author{ Daniel Persson}
\affiliation[Chalmers University of Technology]
{Department of Microtechnology and Nanoscience (MC2), Chalmers University of Technology, SE-412 96 G\"oteborg, Sweden}
\author{Henrik Nilsson}
\affiliation [Lund University]
{Division of Solid State Physics, Lund University, Box 118, S-221 00 Lund, Sweden}
\author{H.Q. Xu}
\affiliation [Lund University]
{Division of Solid State Physics, Lund University, Box 118, S-221 00 Lund, Sweden}
\alsoaffiliation[Peking University]
{Key Laboratory for the Physics and Chemistry of Nanodevices and Department of Electronics, Peking University, Beijing 100871, China}
\author{ Mikael Fogelstr\"om}
\affiliation[Chalmers University of Technology]
{Department of Microtechnology and Nanoscience (MC2), Chalmers University of Technology, SE-412 96 G\"oteborg, Sweden}
\author{ Vitaly Shumeiko}
\affiliation[Chalmers University of Technology]
{Department of Microtechnology and Nanoscience (MC2), Chalmers University of Technology, SE-412 96 G\"oteborg, Sweden}
\author{Per Delsing}
\email{per.delsing@chalmers.se}
\affiliation[Chalmers University of Technology]
{Department of Microtechnology and Nanoscience (MC2), Chalmers University of Technology, SE-412 96 G\"oteborg, Sweden}

\title[Quantized conductance and its correlation to the supercurrent in a nanowire connected to superconductors]
{Quantized conductance and its correlation to the supercurrent in a nanowire connected to superconductors}


\begin{document}

\begin{abstract}
We report conductance and supercurrent of InAs nanowires coupled to Al-superconducting electrodes with short channel lengths and good Ohmic contacts. The nanowires are suspended 15\,nm above a local gate electrode. The charge density in the nanowires can be controlled by a small change in the gate voltage. For large negative gate voltages, the number of conducting channels is reduced gradually and we observe a stepwise decrease of both conductance and critical current before the conductance vanishes completely.

\end{abstract}

The ability to control physical properties and chemical composition of semiconducting nanowires has made them attractive devices to study quantum effects at a very small scale. In particular, semiconducting nanowires coupled to superconductors have been attractive to study quantum transport effects such as quantum interference \cite{Doh}, Cooper-pair field effect \cite{Kewo,  Xu}, and Cooper-pair beam splitting \cite{Schoenenberger}. More recently, they have been used to confirm the presence of exotic states in condensed matter, Majorana Fermion bound states \cite{LeoMaj,LundMaj, Anionic}. 

Semiconducting weak links offer the possibility to gate-control the number of conducting channels, and hence, allow to tune the coupling strength of the weak links. For a ballistic one dimensional weak link, the conductance is expected to increase in quantized steps as a function of gate voltage \cite{Buttiker}. In point contacts, defined in 2DEG heterostructure devices by a split-gate, charge transport can be ballistic and the conductance can be quantized. Evidences of conductance quantization have been realized by many groups in such gate-controlled point contacts \cite{Wees, Wharam}. However, quantization of conductance was only recently reported in nanowires at non-zero magnetic field \cite{Kouwenhoven}. In nanowire weak links this has been difficult mainly due to the scattering of electrons in the conduction paths: reflections of electrons due to scattering centers such as crystal defects, impurities, Schottky barriers, and surface states. Backscattering by such inhomogeneities smears out the conductance steps. In a weak link which is connected to superconducting electrodes there will also be a supercurrent and in line with the conductance quantization, the critical current through such a weak link is also predicted to be quantized with a step height of $\delta I_\ce{c}=e\Delta/ \hbar$ \cite{ Beenakker}. Critical current steps has been previously reported in 2DEG point contacts\cite{Bauch, Takayanagi, Samuelsson} but the observed current steps have been substantially smaller than those predicted by theory. In this paper, we report quantization of conductance in suspended InAs nanowire devices with short channel lengths and good Ohmic contact interfaces. We also observe steps in the critical current which are correlated with the onset of the conductance steps.  

\begin{figure}
\includegraphics[width=0.6\linewidth]{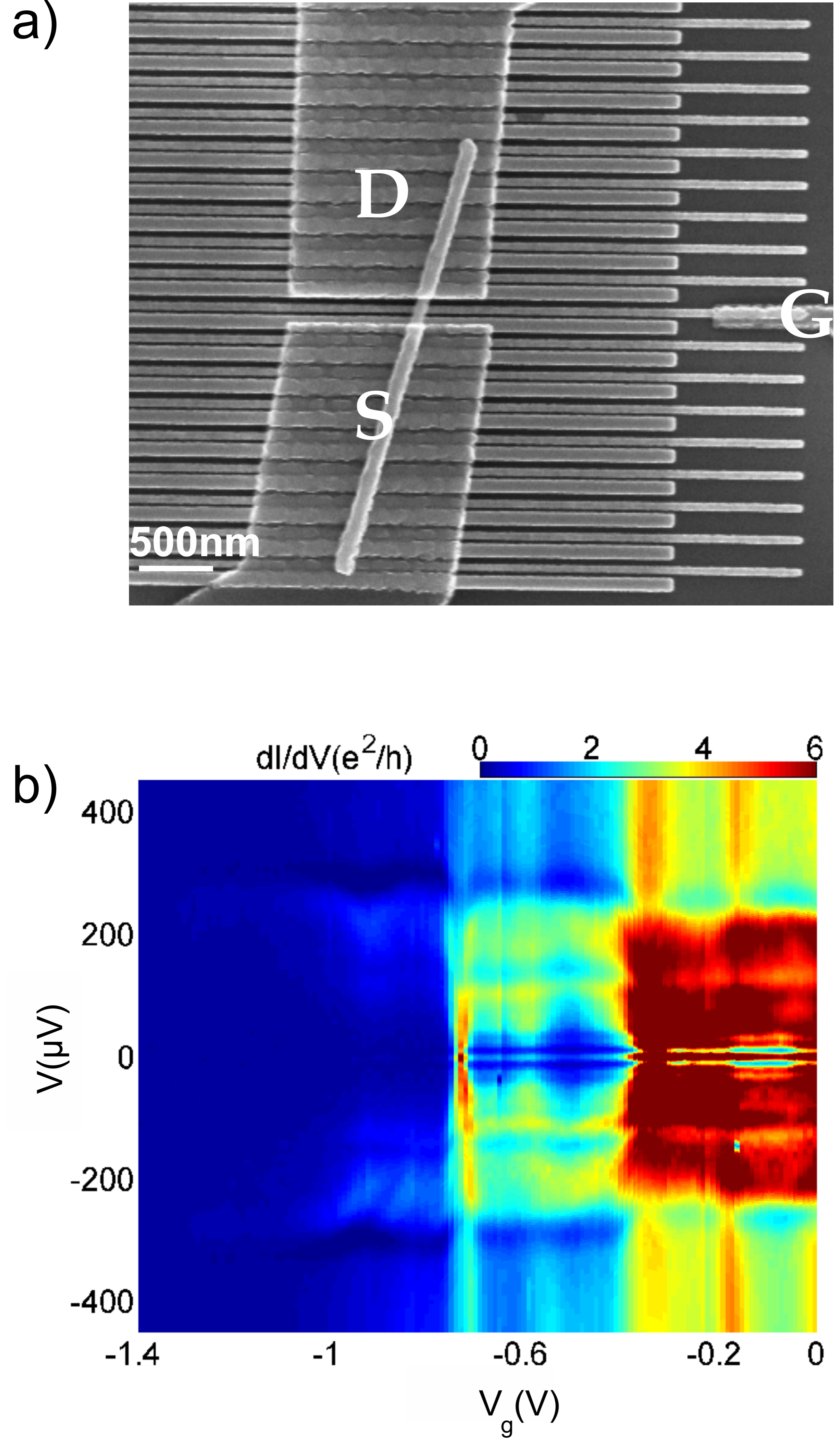}
\caption{ a)  A SEM image of a suspended device. The interdigitated Ti/Au stripes are fabricated with a thickness difference of 15nm such that the nanowire is suspended above the local gate and the substrate. b) An image plot of a differential conductance dI/dV as a function of gate and source-drain voltages. The differential conductance is recorded at the base temperature of 15\,mK. We observe two conductance changes along the gate voltage which mark the opening of the first and the second channels. Similarly, once the first channel starts to conduct, the conductance changes with the source-drain voltage as the transport crosses over from Andreev transport to quasi-particle transport at $V=2\Delta/e\,\approx 250\,\mu$V}
\end{figure}

The geometry of the suspended devices offers the possibility to easily gate-control the number of conducting channels in the nanowires. The control of a few number of conducting channels with nearly ballistic transport will be an important contribution to the ongoing efforts to create topological states, such as Majorana states\cite{LeoMaj,LundMaj,Anionic}.

The nanowires are grown by chemical beam epitaxy\cite{LundFab}. They have an average diameter of 80\,nm and an average length of $4\,\mu$m. To fabricate the suspended devices, a standard Si-substrate capped by 400\,nm thick SiO$_\ce{2}$ is first patterned with interdigitated Ti/Au stripes \cite{Henrik}. The InAs nanowires are then transferred to the already patterned Si substrate and some of the nanowires end up on top of the interdigitated metal stripes. The interdigitated metal stripes are made in a two-step nanofabrication process in order to get a height difference of 15\,nm between every two adjacent stripes. This allows the nanowire to rest on the thicker electrodes while being suspended above the substrate and the thinner electrodes. With the help of scanning electron microscope (SEM) images the positions of suitable nanowires are found and superconducting electrodes (Ti/Al (5/150)\,nm thick) are defined on selected nanowires with electron beam lithography. A SEM image of a suspended device is shown in Fig.\,1a. To get good transparency of the interfaces, an ammonium polysulfide solution (NH$_\ce{4}$$S_\ce{x}$) cleaning process \cite{Lund,Abay} was used prior to evaporation of the superconducting contacts. The samples are then characterized at room temperature and stored in vacuum before further measurements at low temperatures.

Current-voltage characteristics (IVCs) of the devices were measured in a dilution refrigerator with a base temperature of 15\,mK. To decrease noise coupling to the devices, the electrical lines in the measurement set up are well filtered and thermally anchored at different temperature stages of the refrigerator.

\begin{figure}
\includegraphics[width=0.5\linewidth]{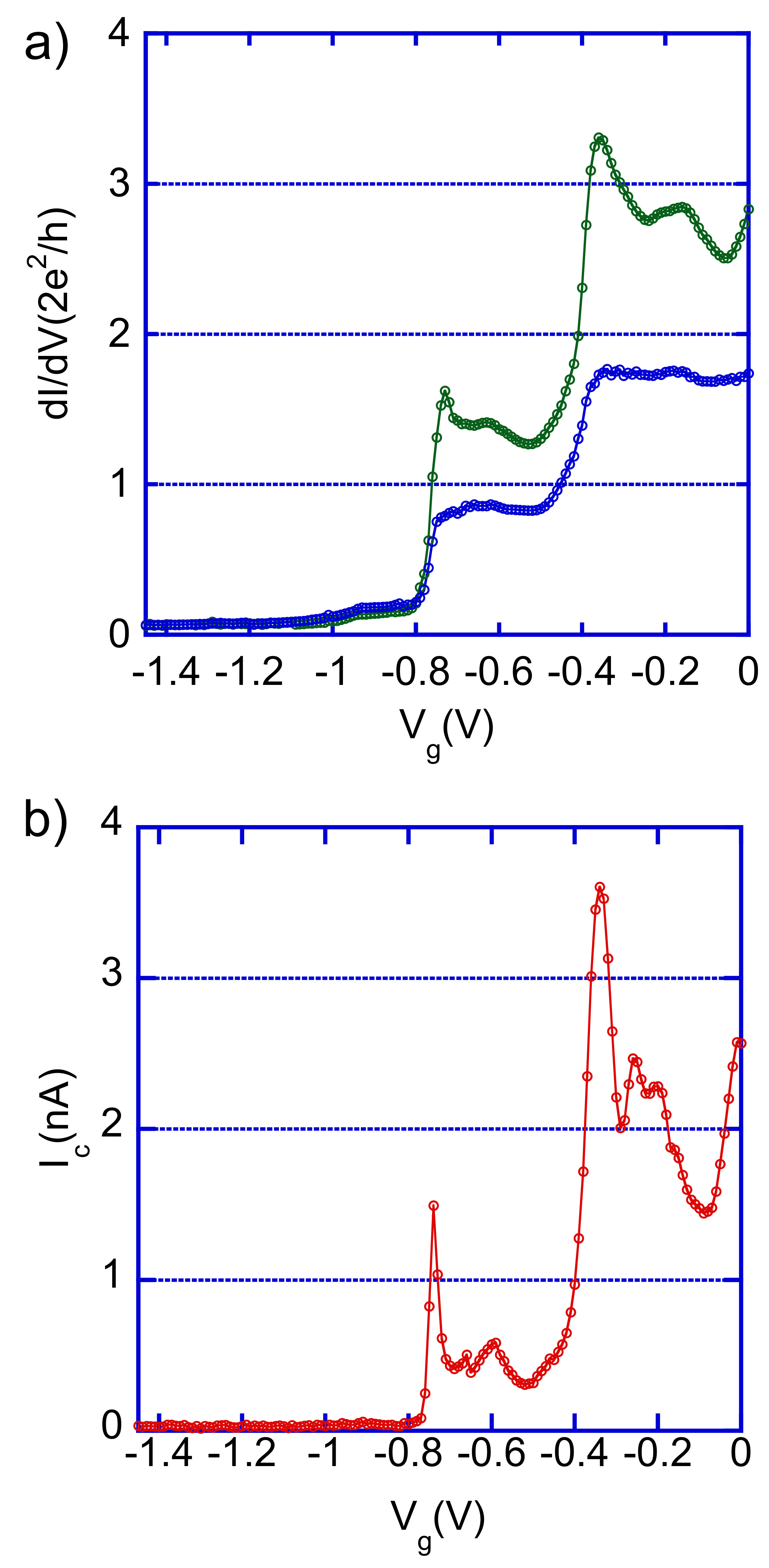}
\caption{ a) Normal state conductance (blue) and a sub-gap conductance (green) as a function of gate voltage. The normal state conductance, extracted at $V =1\,$mV, clearly displays a step wise increase of the conductance as a function of the gate voltage. The sub-gap conductance, taken at $V=125\,\mu$V$\sim\Delta/e$, also increases in steps but with step height larger than the normal state conductance. b) The critical current as a function of the gate voltage. The critical current also increases in steps directly correlated with the normal state conductance.}
\end{figure} 

A typical device with a nearby local gate is shown in Fig.\,1a. We have measured several suspended devices with a broad range of lengths $L$ and normal state resistances $R_\ce{n}$. In general, the devices exhibited intrinsic n-type behavior such that the overall normal state resistance decreased as the gate voltage was increased from negative to positive voltages. The differential conductance dI/dV as a function of gate voltage $V_\ce{g}$ and source-drain voltage $V$  is shown in Fig.\,1b for a junction with $ L\approx$ 150\,nm and $R_\ce{n}=6.7\,k\Omega$ at $V_\ce{g}=0\,V$. We observe transition from tunneling regime to conducting regime and stepwise increases of the conductance due to the opening of conducting channels. These conductance changes mark the opening  of the first and the second channels with increasing gate voltage. Similarly, along the vertical axis, in the conducting regime, we observe a major conductance change at the voltage of $|2\Delta/e|\sim\,250\,\mu$V where the transport crosses over from Andreev transport at low bias to single-particle transport at high bias.

To clearly display the observed conductance changes as a function of the gate voltage, two line cuts of the differential conductance are shown in Fig.\,2a. The line cuts correspond to two source-drain voltages, at large bias, $V =$1m$\,$V $>> 2\Delta$/e (blue) where the conductance corresponds to the normal state value, and  the sub-gap state $V= 125\,\mu$V $ \sim\Delta/e$ (green). The normal state differential conductance dI/dV at $V >> 2\Delta$/e displays clear quantization of the conductance as a function of the gate voltage. As the gate voltage is increased to more positive values, the potential barrier height decreases and the channels become conducting one by one. The measured step heights are almost equal to one quantum of conductance $2e^{2}/h$ owing to the short channel lengths and indicating almost perfect transmission of the conducting channels. The intrinsic interface resistance due to the Fermi-velocity mismatch and the non-zero resistance of the Ohmic contacts explain the small deviations in the step height from the ideal quantum of conductance.

The sub-gap conductance also increases step wise as a function of the gate voltage. A differential conductance at a source-drain voltage of $V = 125\,\mu$V $ \sim\Delta/e$ (green) is shown in Fig.\,2a. The measured step height is bigger than one quantum of conductance ($2e^{2}/h$) which is attributed to the two-particle charge transport mechanism of Andreev reflection\cite{AndreevT, BTK}.

Next, we present the critical current as a function of the gate voltage, as shown in Fig.\,2b.  At high negative gate voltages, the critical current is completely suppressed in the same way as the normal state conductance. When the gate voltage reaches $V_\ce{g}\approx-0.8\,$V, at which the first channel starts to conduct, the critical current starts to appear. It shows a peak of 1.5\,nA at the channel onset, but then rapidly decreases to a plateau with an average of 0.5\,nA. When the gate voltage was further increased to open a second conducting channel, the critical current again shows a peak (to 3.6\,nA) and then settles down to an average of 2\,nA. The step wise increase of the critical current is directly correlated to the step wise increase of the normal state conductance. However, the observed step height is substantially lower than the maximum theoretical value $I_\ce{c}(T)=\dfrac{e\Delta}{\hbar}(1-\sqrt{1-T})\approx14$\,nA,\cite{Chtchelkatchev} where the transmission probability of the first channel $T \approx 0.8$ is estimated from the step height of the  normal state conductance. The reason for the small critical current is not obvious but it might be a result of early switching of the critical current due to thermal activation. We note that this discrepancy is however not only unique to our device, this has also been the case in previous reports. \cite{Bauch,Takayanagi, Samuelsson}. 

\begin{figure}
\includegraphics[width=0.6\linewidth]{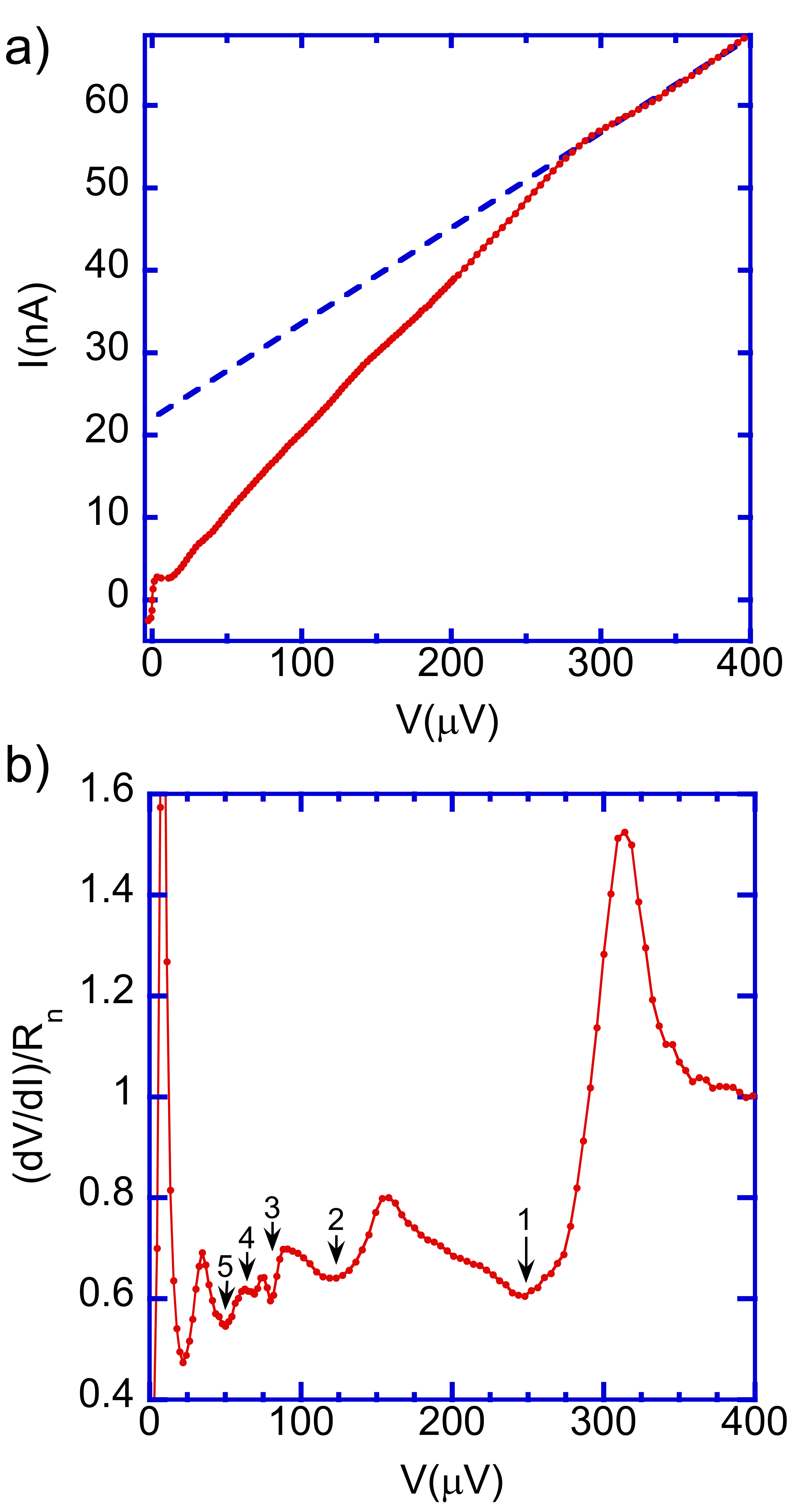}
\caption{ a)  Current voltage characteristic taken at $ V_{g}=0$\,V and at a temperature of 15\,mK.  Above $V > 2\Delta/e$, the current voltage characteristic shows a linear behavior with a normal state resistance of $R_\ce{n}=6.7\,k\Omega$. Once $V \leq2\Delta/e$, the current voltage characteristic shows an enhanced conductance evident by a change of slope (visible as a kink). The excess current is shown with a blue dashed line, a linear fit of the data in the normal state extrapolated to $V=0$. b) Normalized differential resistance $(dV/dI)R_n$ of the current-voltage characteristics in (a). The sub-gap resistance changes to almost half of the normal state resistance. Sub-gap features from multiple Andreev reflections are observed at $2\Delta/n$e, where $n=1,2,3,4$, and 5.}
\end{figure}

To observe the conductance quantization and the correlated critical current steps, high quality interfaces are necessary. Here, we extract an additional information about the contact interface transparency by quantifying the excess current \cite{BTK}. We have plotted IVC of the device in Fig.\,3a. The IVC is taken at $V_\ce{g}=0\,V$ which corresponds to two open conduction channels. At high voltages, $V > |2\Delta$/e$|\,\sim250\mu V$, the IVC displays a linear behavior with a slope of the normal state conductance. Once the voltage reaches $V \sim2\Delta$/e, we observe enhancement of the conductance evident by the change of slope in the IVC. A linear fit (blue dashed line) to the normal state (above $2\Delta$) does not extrapolate to the origin but to an excess current, $I_\ce{exc}\approx23$\,nA. The interface transparency $T$  can be extracted from the extrapolated excess current. Using the formula for the excess current derived in  Ref.\,\cite{Shumeiko}), we calculated the contact average transparency $T=0.67$, for $\Delta\sim125\,\mu eV$, and $ R_\ce{n}=6.7\,k\Omega$. The extracted value is of the same order as the one for the first step, and it is close to the transmission probability extracted from the normal resistance.

In order to clearly display the enhanced sub-gap conductance and the sub-gap structures, we have presented normalized differential resistance $dV/dI$ as a function of source-drain voltage for $V_\ce{g}=0\,V$. As can be seen in Fig.\,3b, the resistance drops substantially when the device goes from quasi-particle transport to Andreev transport. The normal state resistance at $V >> |2\Delta/e|$, is almost twice the sub-gap resistance at $V << |2\Delta/e|$ (cf. experimental data for InAs 2DEG in Ref.\cite{Samuelsson}\, Fig 2, and theoretical calculation in Ref.\,\cite{Cuevas}\,  Fig 3). The plot also clearly presents anomalous multiple Andreev reflection (MAR) features at voltage positions corresponding to $(2\Delta/ne)$. The first five, $n= 1, 2, 3,4, 5$  indicated by arrows in Fig.\,3b, scales with the superconducting gap. These results are similar to the differential resistance dV/dI of quasi-ballistic InAs (2DEG) devices\cite{Samuelsson}. However, the pronounced peak, near to zero bias voltage does not depend on the superconducting gap and we therefore believe that its origin is different from MAR.

In conclusion, we have fabricated devices of InAs nanowires suspended above a local gate electrode with short channel lengths and good Ohmic contacts. The local charge concentration and hence the conductance of the nanowires was quite sensitive to small changes of the gate voltage. As a function of the gate voltage, the normal state conductance increased in steps with step heights almost one conductance quantum $2e^{2}/h$. Similarly, the sub-gap conductance and the critical current also increased in steps directly correlated with the normal state conductance.

\acknowledgement

We acknowledge fruitful discussions with Lars Samuelson, Christopher Wilson and Jonas Bylander. The work was supported by the Swedish Research council, the Wallenberg foundation, and the National Basic Research Program of the Ministry of Science and Technology of China (Nos. 2012CB932703 and 2012CB932700).

%\bibliography{achemso}

\end{document}